\documentclass[12pt]{iopart}

\usepackage{epsfig,subfigure}
\usepackage{rotating}

\newcommand{\msun}{M_\odot}
\newcommand{\be}{\begin{equation}}
\newcommand{\ee}{\end{equation}}
\newcommand{\bea}{\begin{eqnarray}}
\newcommand{\eea}{\end{eqnarray}}
\newcommand{\prd}{{\it Phys. Rev.} D }
\newcommand{\cqg}{\it Class. Quantum Grav. }

\begin{document}

\title[Application of the effective Fisher matrix]{Application of the effective Fisher matrix to the frequency domain inspiral waveforms}

\author{Hee-Suk Cho and Chang-Hwan Lee}
\address{Department of Physics, Pusan National University, Busan 609-735, Korea}
\ead{chohs1439@pusan.ac.kr}

\begin{abstract}
The Fisher matrix (FM) has been generally used to predict the accuracy of the gravitational wave parameter estimation. Although the limitation of the FM has been well known, it is still  mainly used due to its very low computational cost compared to the Monte Carlo simulations. Recently, Rodriguez {\it et al.} [Phys. Rev. D {\bf 88}, 084013 (2013)] performed  Markov chain Monte Carlo (MCMC) simulations using a frequency domain inspiral waveform model (TaylorF2) for nonspinning binary systems with total masses   $M \leq 20 \msun$,  they found  systematic differences between the predictions from FM and MCMC for $M>10 \msun$. On the other hand, an effective Fisher matrix (eFM)  was recently introduced by Cho {\it et al.} [Phys. Rev. D {\bf 87}, 24004 (2013)].
The eFM is a semi-analytic approach to the standard FM, in which the derivative is taken of a quadratic function fitted to the local overlap surface.
In this work, we apply the eFM method to the TaylorF2 waveform for nonspinning binary systems with a moderately high signal to noise ratio (${\rm SNR} \sim 15$) and find that the eFM can well reproduce the MCMC error bounds in Rodriguez {\it et al.} even for high masses.
By comparing the eFM standard deviation directly with the 1-$\sigma$ confidence interval of the marginalized overlap that approximates the MCMC posterior distribution, we show that the eFM can be acceptable in all mass regions for the estimation of the MCMC error bounds.
We also investigate the dependence on the signal strength.
\newline
\newline
\noindent{Keywords: gravitational waves, compact binary coalescence, parameter estimation, Fisher matrix}
\end{abstract}


\section{Introduction}
In ground-based gravitational wave data analysis, various parameter estimation methods are implemented to find the physical parameters of the gravitational wave sources.
One of the most promising techniques is the Markov chain Monte Carlo (MCMC)~\cite{Cor06,Van09,Cor11,Vei12,Osh141,Osh142}, which involves the Bayesian analysis framework.
The MCMC enables us to search the whole parameter space within given templates and find the physical parameters from the wave signal and the error bounds on their variances.
The Fisher matrix (FM) has been generally used to estimate the error bounds~\cite{Poi95,Aru05,Lan06,Bro07,Wad13,Nie13}.
Despite the well-known limitations~\cite{Val08}, the FM method is being used because it is quite easy to use and needs very low computational cost compared to the Monte Carlo simulations.
Though most of real waveforms may extend continuously to the merger and ringdown phases, the frequency domain inspiral-only waveform model, so called ``TaylorF2", have been generally considered in the past FM studies because the waveform can be given by an analytic function so that it is very easy to calculate the derivatives in the FM formalism.

Meanwhile, several studies investigated the inconsistencies between the FM and Monte Carlo methods~\cite{Cor06,Cok08,Rod13} for the TaylorF2 model, especially, Rodriguez {\it et al.}~\cite{Rod13} (henceforth denoted RFFM) performed a systematic comparison between the FM error estimates and the MCMC probability density functions using plenty of nonspinning binary systems. They found that the FM overestimates the uncertainty in the parameter estimation achievable by the MCMC in high mass region ($m_1+m_2=M > 10\msun$), and the disagreement increases with the total mass.
The inconsistency between the FM and the MCMC  was first noted by Cornish {\it et al.}~\cite{Cor06}, and robustly confirmed by RFFM.
They explored various possibilities but could not find a convincing explanation on the discrepancy.
Most recently, Mandel {\it et al.}~\cite{Man14} performed in-depth study of this issue and clearly explained the origin of the discrepancy

In the TaylorF2 model, the waveform is abruptly terminated at a certain cutoff frequency ($f_{\rm cut}$) which depends on the binary's total mass.
However, the FM formalism assumes that all the fiducial parameters of a signal are known.
Hence, $f_{\rm cut}$ of the comparing template is set to be the same as the signal's cutoff frequency, and has been ignored in the past FM studies. Moreover, when considering the mass dependent $f_{\rm cut}$, the analytic FM cannot be determined because the step function introduced by $f_{\rm cut}$ is not differentiable.
In the MCMC formalism, on the contrary, $f_{\rm cut}$ should be continuously varying with the template mass because the real parameter values of the signal are hidden.
Mandel {\it et al.}~\cite{Man14} showed that the omission of the (template) mass dependent $f_{\rm cut}$ in the FM induces the discrepancy, and the disagreement between the FM and the MCMC can increase for higher masses as shown in RFFM.

On the other hand, an effective Fisher matrix (eFM) was recently developed by Cho  {\it et al.}~\cite{Cho13}.
They showed a good agreement between the eFM and MCMC error predictions on the mass parameters using a time domain inspiral waveform for a black hole ($10 \msun$)-neutron star ($1.4 \msun$) binary~\cite{Osh141,Osh142}.
While the FM computation is based on the differentiation of a wave function, the eFM  is based on the derivatives of a quadratic fitting function which is fitted to the local region of the MCMC posterior distribution and always differentiable. Therefore, if the local posterior surface can be approximated by a quadratic function, the eFM will give a good estimate on the MCMC probability density function.

In this work, we apply the eFM method to the TaylorF2 model for nonspinning binary systems with the same mass ranges as in RFFM.
By comparing the eFM results with the MCMC posteriors, we show that the eFM method is acceptable for the estimation of the MCMC error bounds for all mass ranges, and the accuracy is substantially improved compared to the standard FM at the high mass region.
In section \ref{sec.wavefunction}, we outline the TaylorF2 waveform model and the FM formalism.
In section \ref{sec.likelihood}, we investigate a property of the MCMC likelihood derived from the overlap surface~\cite{Cho13} and provide concrete examples showing the difference between the FM and MCMC overlap formalisms. In section \ref{sec.effectivefisher}, we briefly review the eFM method and discuss the validity of this effective machinery. The results are summarized in section \ref{sec.result}. We compute the standard deviations using the eFM method and compare with those of the FM.
Then, we demonstrate that the divergent trend in fractional differences is consistent with that of RFFM.
In particular, giving a direct comparison between the standard deviation of the eFM and the 1-$\sigma$ confidence interval of the overlap,
we provide valid criteria of the FM and eFM to estimate the MCMC error bounds for the TaylorF2 model.
We also discuss the dependence on the signal strength as well as a limitation of the MCMC formalism for the inspiral-only waveforms.
Finally, we conclude this work by summarizing the result in section \ref{sec.conclusion}.


\section{Wave function and Fisher matrix}\label{sec.wavefunction}
We use the TaylorF2 waveform that is implemented in the LIGO Algorithm Library~\cite{lal}.
The analytic function of TaylorF2 waveform is given by
\be
\tilde{h}(f)=Af^{-7/6}e^{i\Psi(f)},
\ee
where $A$ is the wave amplitude that consists of the binary masses and five extrinsic parameters, i.e., the luminosity distance of the binary, two angles defining the sky position of the binary system with respect to the detector, the orbital inclination ($\iota$), and the wave polarization ($\psi$).
For simplicity, in our calculation, only a single detector configuration is adopted and the extrinsic parameters are not considered. Note that we assume a fixed signal-to-noise ratio (SNR) and the phase rather than the amplitude is the main determining factor in the mismatch calculation. 
All information of the waveform is determined by the post-Newtonian (pN) phase function $\Psi(f)$. The coefficients of $\Psi(f)$ consist of  the chirp mass ($M_c=m_1^{3/5}m_2^{3/5}M^{-1/5}$), symmetric mass ratio ($\eta=m_1m_2M^{-2}$), coalescence time ($t_c$), and termination phase ($\phi_0$):
\be\label{eq.phasing}
\Psi(f)=2\pi f t_c - 2 \phi_0 - {\pi \over 4} + {3 \over 128 \eta}\phi(M_c,\eta, f),
\ee
where  $t_c$ can be chosen arbitrarily,  and $\phi(M_c,\eta, f)$ can be represented by the pN expansion, 
in this work, we consider up to 3.5 pN order~\cite{Aru05}.

The termination phase ($\phi_0$) is related to the coalescence phase ($\phi_c$) by~\cite{Sat91, All12}
\be \label{eq.phi0}
2\phi_0=2\phi_c-{\rm arctan}\bigg({F_{\times} \over F_+ }{2 \cos \iota \over 1+\cos^2\iota}\bigg),
\ee
where $F_{\times}$ and $F_+$ are the antenna response functions depending on $\psi$ and the sky position.
For a fixed binary system, $\phi_0$ is a function of $\iota, \psi$, and the coalescence phase $\phi_c$ (the coalescence phase can also be chosen arbitrarily). 
In the past works (e.g., \cite{Poi95,Cut94}), $\phi_0$ has been generally assumed to be an arbitrary constant in FM computations. However, in order to take into account more than two angle parameters among ($\iota, \psi, \phi_c$), one should define the $\phi_0$ as a function of the angle parameters ($\iota, \psi, \phi_c$). For example, if the binary is optimally placed and orientated, then $\phi_0=\phi_c - \psi$. In this case, the FM is singular and the inverse matrix cannot be defined. The correlation between these two parameters becomes reduced as the $\iota$ increases.
If $\iota=\pi/2$, $\phi_0$ is equal to the arbitrary constant $\phi_c$ and other angle parameters can be removed from the phase equation.
In this work, since we do not consider the extrinsic parameters, $\phi_0$ is assumed to be the same as in the previous works, then the wave phase in Eq.~(\ref{eq.phasing}) is determined by a combination of the parameters ($M_c, \eta, t_c, \phi_c$), and we only consider these four parameters in the FM.
Note that, however, when computing the analytic FM which includes both $\phi_c$ and $\psi$, one should not set $\phi_0$ equal to $\phi_c$ in general.

The high frequency cutoff ($f_{\rm cut}$) of the TaylorF2 wave function is taken when the binary hits the ``innermost- stable-circular orbit (ISCO)", that is defined as a function of the total mass ($M$) of the system:
\be \label{eq.flso}
f_{\rm cut}=f_{\rm ISCO}={1 \over 6^{3/2}\pi M},
\ee
and the low frequency cutoff ($f_{\rm min}$) is fixed to be 40 Hz independently of the mass.
Then, the overlap (match) between a signal ($\tilde{h}_s$) and a template ($\tilde{h}_t$) is defined by
\be \label{eq.conventionaloverlap}
\langle \tilde{h}_s | \tilde{h}_t \rangle =  4 {\rm Re} \int_{f_{\rm min}}^{f_{\rm cut}}  \frac{\tilde{h}_s(f)\tilde{h}^*_t(f)}{S_n(f)} df,
\ee
where $S_n(f)$ is a detector noise power spectrum, we adopt a model for the initial LIGO~\cite{Dam01} to provide concrete testbed results.
Note that the inverse Fourier transform will compute the overlap for all possible coalescence times at once~\cite{All12}. In addition, by taking the absolute value of the complex number we can maximize the overlap over all possible coalescence phases~\cite{All12},
Here, as noted in~\cite{Aji09}, one should be confident that the true maximum is never missed in the sufficiently small tolerance level for the maximizing algorithm.
To do this, we apply a nearly continuous time shift by reducing a step size\footnote{This can be done by zero padding in the frequency domain data to lower the time domain sample spacing.} when performing the inverse fast Fourier transform.
Finally, we define the normalized overlap by
\be\label{eq.maxoverlap}
P(\tilde{h}_s,\tilde{h}_t) =  {\rm max}_{t_{\rm c},\phi_c}{\langle \tilde{h}_s | \tilde{h}_t \rangle   \over  \sqrt{\langle \tilde{h}_s | \tilde{h}_s \rangle \langle \tilde{h}_t | \tilde{h}_t \rangle }} .
\ee

The FM for a waveform $\tilde{h}(\lambda)$ is defined by
\be \label{eq.analyticfisher}
\Gamma_{ij}=\bigg\langle {\partial \tilde{h} \over \partial \lambda_i} \bigg | {\partial \tilde{h} \over \partial \lambda_j} \bigg \rangle\bigg|_{\lambda=\lambda_0},
\ee
where $\lambda_0$ is the true value of each parameter and $\lambda_i=\{M_c, \eta, \phi_c, t_c\}$. Since the pN phase of the TaylorF2 is an analytic function of the parameters as in equation~(\ref{eq.phasing}), the derivatives can be obtained analytically. 
And the overlap integration is performed in $[f_{\rm min}, f_{\rm cut}]$ as in equation~(\ref{eq.conventionaloverlap}).
On the other hand, the FM can be directly derived from the log likelihood ($\ln L$)~\cite{Jar94,Val08} and the loglikelihodd can be expressed approximately by~\cite{Cho13}
\be \label{eq.LvsP}
\ln L(\lambda) =-\rho^2  (1-P),
\ee
where $\rho$ is the SNR and $P$ is the normalized overlap in equation~(\ref{eq.maxoverlap}).
Therefore, we have
\be \label{eq.numericalfisher}
\Gamma_{ij}=-{\partial^2 \ln L(\lambda) \over \partial \lambda_{i} \partial \lambda_{j}}\bigg|_{\lambda=\lambda_0}=-\rho^2  {\partial^2 P(\lambda) \over \partial \lambda_{i} \partial \lambda_{j}}\bigg|_{\lambda=\lambda_0}.
\ee
The overlap $P$ is obtained after maximizing the inner product of the unit-norm signal and the unit-norm template over $t_c$ and $\phi_c$. 
In this work, we assume a uniform prior for all parameters, and the likelihood can be a Gaussian distribution in the limit of high SNR.
Then, the process of marginalizing a likelihood over $t_c$ and $\phi_c$ is formally equivalent to maximizing the likelihood in those parameters, up to an irrelevant normalization constant~\cite{Cho13}. Therefore, $\Gamma_{ij}$ corresponds to a marginalized $2 \times 2$ FM consisting of $\lambda_i=\{M_c, \eta\}$.

\section{Likelihood}\label{sec.likelihood}

In equation~(\ref{eq.numericalfisher}), we find that the FM can be calculated by differentiating the overlap surface, and this overlap depends on the frequency cutoff $f_{\rm cut}$ in the integration in equation~(\ref{eq.conventionaloverlap})  when the TaylorF2 model is used.
In the MCMC formalism, $f_{\rm cut}$ of the template is continuously varying with the total mass.
On the other hand, $f_{\rm cut}$ is the same for both the signal and the template in the FM formalism.
To test the difference between the MCMC and FM overlaps, we briefly show two different results for a given waveform set, $h_s$  and $h_t$ (for a detailed analytic approach, see \cite{Man14}).
We denote the frequency cutoffs of the $h_t$ and $h_s$ by $f^t_{\rm cut}$ and $f^s_{\rm cut}$, respectively.

\begin{figure}[t]
\includegraphics[width=\columnwidth]{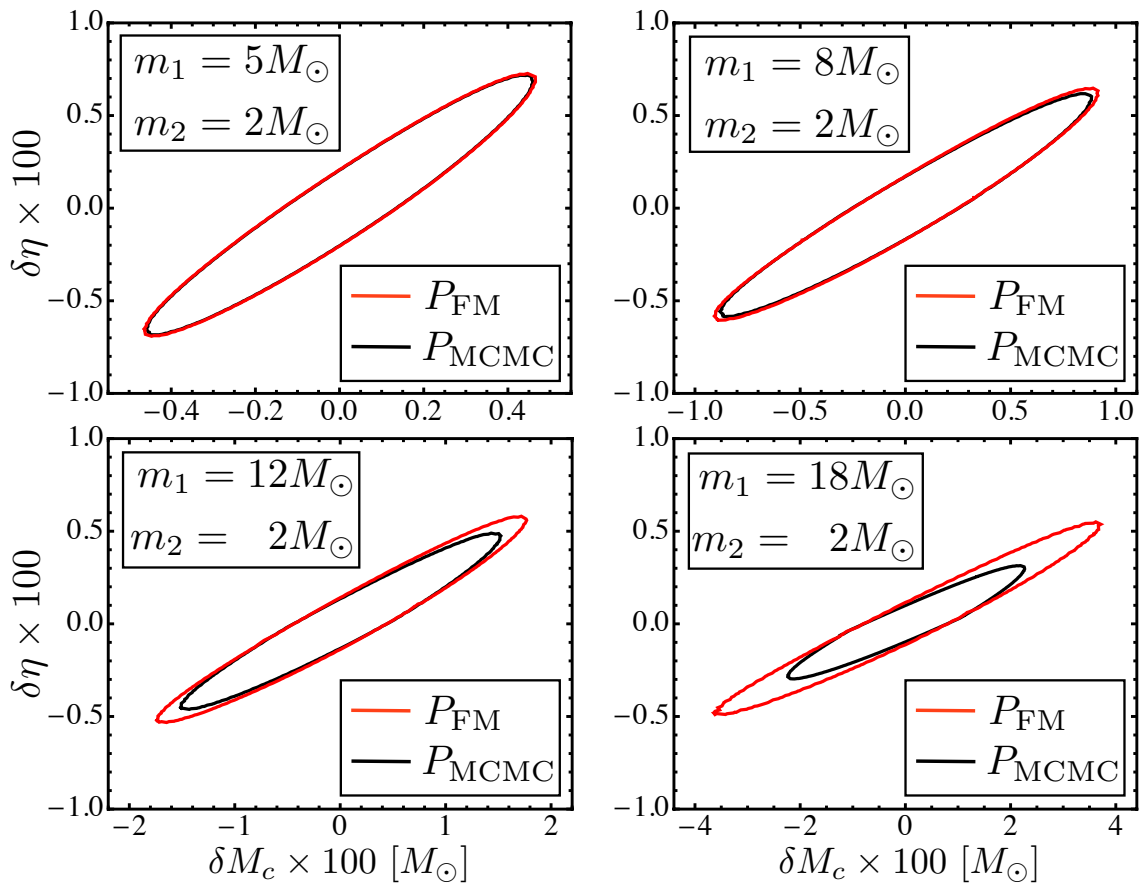}
\caption{\label{fig1}Comparison between the MCMC and FM overlaps with various total masses. The contours indicate $P=0.995$. $P_{\rm FM}$ and $P_{\rm MCMC}$ are calculated using the FM [equation~(\ref{eq.independentP})] and MCMC [ equation~(\ref{eq.dependentP})] overlap formalisms, respectively. The component masses are denoted in each plot. Note that the difference between two overlap contours is more distinguishable as the total mass increases.}
\end{figure}

\begin{itemize}

\item[(i)]
{\bf FM formalism} : \\
In the FM formalism, $f^t_{\rm cut}=f^s_{\rm cut}$ is assumed independently of the total mass of the template. We reexpress the  overlap $P$ by combining equations~(\ref{eq.conventionaloverlap}) and (\ref{eq.maxoverlap}) by
\be\label{eq.independentP}
P_{\rm FM} ={\rm max}_{t_{\rm c},\phi_c} {\langle \tilde{h}_s | \tilde{h}_t \rangle |_{f_{\rm min}}^{f^s_{\rm cut}}   \over  \sqrt{\langle \tilde{h}_s | \tilde{h}_s \rangle|_{f_{\rm min}}^{f^s_{\rm cut}} \langle \tilde{h}_t | \tilde{h}_t \rangle|_{f_{\rm min}}^{f^s_{\rm cut}} }},
\ee
where ``$|_a^b$" means that the overlap integration should be performed only in the frequency range from $a$ to $b$.

\item[(ii)]
{\bf MCMC formalism} :  \\
In the MCMC formalism, $f_{\rm cut}^t$ is not fixed but dependent on the template mass, so generally $f_{\rm cut}^t \neq f_{\rm cut}^s$.
Then, the overlap, which contributes the MCMC overlap formalism, should be~\cite{Ohm13}
\be\label{eq.dependentP}
P_{\rm MCMC} = {\langle \tilde{h}_s | \tilde{h}_t \rangle |_{f_{\rm min}}^{{\rm min} (f^t_{\rm cut},f^s_{\rm cut})}   \over  \sqrt{\langle \tilde{h}_s | \tilde{h}_s \rangle|_{f_{\rm min}}^{f^s_{\rm cut}} \langle \tilde{h}_t | \tilde{h}_t \rangle|_{f_{\rm min}}^{f^t_{\rm cut}} }},
\ee
where ``min($a,b$)" indicates the smaller value among $a$ and $b$.
In this case, $\langle \tilde{h}_s | \tilde{h}_s \rangle$ is independent of the template, hence always the same as that of $P_{\rm FM}$, however, the others can be different.
First, if $f^t_{\rm cut}>f^s_{\rm cut}$, then  ${\rm min} (f^t_{\rm cut},f^s_{\rm cut})=f^s_{\rm cut}$ and the numerator is the same as that of $P_{\rm FM}$, while
$\langle \tilde{h}_t | \tilde{h}_t \rangle|_{f_{\rm min}}^{f^t_{\rm cut}}>\langle \tilde{h}_t | \tilde{h}_t \rangle|_{f_{\rm min}}^{f^s_{\rm cut}}$ due to the integral fraction $\langle \tilde{h}_t | \tilde{h}_t \rangle|_{f^s_{\rm cut}}^{f^t_{\rm cut}}$, so the denominator should be larger than that of $P_{\rm FM}$. Therefore, we have $P_{\rm MCMC}<P_{\rm FM}$.
Second, if $f^t_{\rm cut}<f^s_{\rm cut}$, then ${\rm min} (f^t_{\rm cut},f^s_{\rm cut})=f^t_{\rm cut}$ and both $\langle \tilde{h}_s | \tilde{h}_t \rangle$ and $\langle \tilde{h}_t | \tilde{h}_t \rangle$ are smaller than those of $P_{\rm FM}$.
In this case, the numerator decreases more rapidly than the denominator, and we also find $P_{\rm MCMC}<P_{\rm FM}$.

\end{itemize}

There exists non vanishing difference between the two overlaps ($\Delta P=P_{\rm FM}-P_{\rm MCMC}$) when $f^t_{\rm cut} \neq f^s_{\rm cut}$,
which turns out to be nonnegligible in the high mass region.
To see the dependence of $\Delta P$ on the mass, we show the overlap contours for both $P_{\rm FM}$ and $P_{\rm MCMC}$ with various total masses in figure~\ref{fig1}.
One can see that the overlap contours are narrower for the MCMC formalism and the difference between the MCMC and FM overlaps is more distinguishable for more massive total masses.
As we will see in the next section, the overlap contours are directly related with the confidence region of the MCMC posterior, so
the narrower ellipsoids correspond to the smaller error bounds in parameter estimation.  Therefore, this result briefly shows the trend of discrepancy between the FM and MCMC error predictions summarized in RFFM. 


\section{Validity of the effective method}\label{sec.effectivefisher}

In equation~(\ref{eq.numericalfisher}), although the wave function is an analytic equation, the overlap surface $P$ cannot be expressed by an analytic function, hence we have to calculate the partial derivatives numerically.
However, since the local region of the overlap is generally a quadratic distribution (this is because the likelihood can be a Gaussian distribution in the limit of high SNR), a multivariate quadratic function fits the local overlap surface reasonably good enough. 
So if we find an analytic fitting function $F$ to the overlap $P$ at a certain fitting region above $P_{\rm min}$  (i.e., $P \geq P_{\rm min}$), the derivatives can be analytically obtained.
Using this function, Cho {\it et al.}~\cite{Cho13} defined the eFM by
\be
(\Gamma_{\rm eff})_{ij}=-\rho^2 {\partial^2 F(\lambda) \over \partial \lambda_{i} \partial \lambda_{j}}\bigg|_{\lambda=\lambda_0}.
\ee

The fitting region is roughly related with the SNR as $P_{\rm min} \sim 1-1/ \rho^2$ in the eFM approach~\cite{Cho13}.
A general expression that incorporates a dependence on the number of parameters has been derived by Baird {\it et al.}~\cite{Bai13},
where the confidence region of the MCMC posterior can be directly approximated by the overlap as
\be \label{eq.P}
P\geq1-{\chi^2_k(1-p) \over 2 \rho^2 },
\ee
where $\chi^2_k(1-p)$ is the chi-square value for which there is $1-p$ probability of obtaining that value or larger and the $k$ denotes the degree of freedom,  given by the number of parameters.
Since we consider the two-dimensional overlap surfaces (i.e., $k=2$), the 1-$\sigma$ confidence region (i.e., $p=0.68$) at a given SNR is given by
\be
P \geq 1-{1.14 \over \rho^2}. \label{eq.P2d}
\ee
If we assume the SNR to be 15, the confidence region can be $P \geq 0.995$.
Finally, we choose this overlap region for the fitting function, therefore, our fitting region is physically motivated by the moderately high SNR.

For the given fitting region ($P_{\rm min}=0.995$), we can test the accuracy of the effective machinery by comparing to the analytic method.
Recall that the overlap surface $P_{\rm FM}$ in figure~\ref{fig1} is computed by the standard (analytic) FM formalism.
Hence, if we obtain the eFM by using the fitting function to the overlap $P_{\rm FM}$,
the result should be consistent with that of the analytic FM.
We summarize the comparison results in table~\ref{tab.fishercomparison} for a low mass and a high mass binary models.
As discussed above, the resultant eFM will be the $2 \times 2$ matrix marginalized over $t_c$ and $\phi_c$, so we only consider two mass parameters $M_c$ and $\eta$.
For the binary models given, we find a very good agreement between the two methods within 1 $\%$\footnote{The origin of a tiny discrepancy is because the overlap surface is not perfectly quadratic at the given fitting scale. This discrepancy can be reduced by choosing smaller fitting scales.}, this indicates that the marginalization algorithm in overlap computation is accurate and the effective method can reproduce the analytic result exactly. 
In this table, note that, we use the overlap $P_{\rm FM}$ for the eFM.
In the next section, we will apply the eFM method to the overlap $P_{\rm MCMC}$ to reproduce the MCMC error bounds.

\begin{table}
\caption{\label{tab.fishercomparison}{Comparison between the effective and analytic FM methods using the overlap surface $P_{\rm FM}$. We assume the SNR of 15, so the fitting region of  $P_{\rm min}=0.995$. The correlation coefficient is denoted by $c_{ij}$. The consistency between the analytic and effective methods indicates that  the marginalization algorithm in overlap computation is accurate and the effective method can reproduce the analytic result exactly.}}
\begin{indented}
\item[]\begin{tabular}{c  ccccccccc  }
\br
$m_1, m_2$                                    &                                      & \multicolumn{4}{c}{$5 \msun$, $2 \msun$}       &  \multicolumn{4}{c}{$18 \msun$, $2 \msun$}    \\
\mr
 Method       &&\multicolumn{2}{c}{Analytic}  &\multicolumn{2}{c}{Effective}  &\multicolumn{2}{c}{Analytic}  &\multicolumn{2}{c}{Effective} \\
\mr
 Parameter&&$M_c [\msun]$ & $\eta$  &     $M_c [\msun]$   &  $\eta$      &$M_c [\msun]$ &  $\eta$    &$M_c [\msun]$ &  $\eta$\\ 
\mr
 $\sigma_i \times 10^3$		           & &3.10  &4.73      &3.07  &4.72&25.1  &3.54&24.9  &3.56\\
\mr
 $c_{ij}$        &&\multicolumn{2}{c}{0.957}  &\multicolumn{2}{c}{0.957}  &\multicolumn{2}{c}{0.976}  &\multicolumn{2}{c}{0.976} \\
\br
 \end{tabular}
 \end{indented}

\end{table}

\section{Result}\label{sec.result}


\subsection{Indirect comparison to the MCMC}
\begin{figure}[t]
\includegraphics[width=8cm]{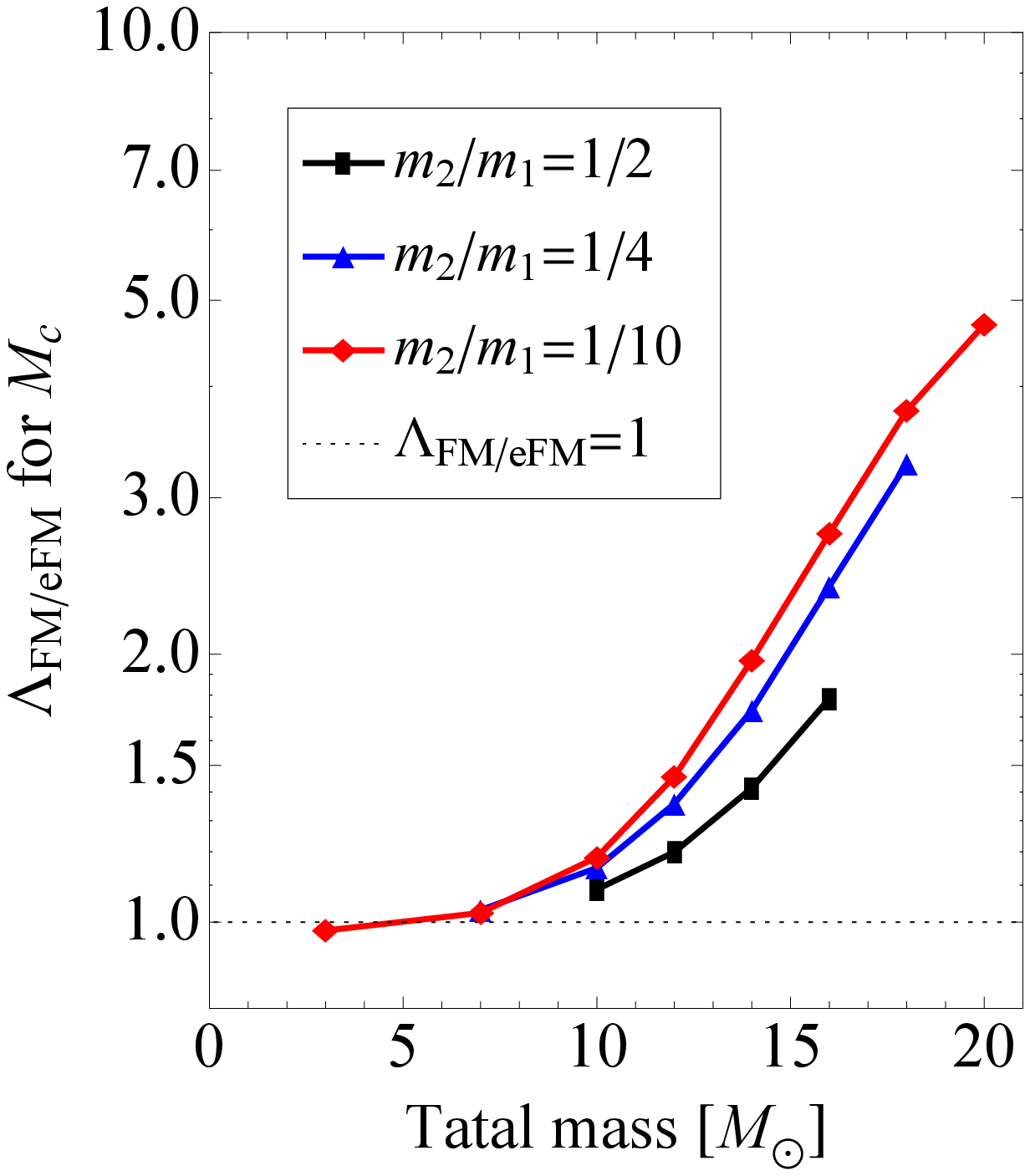}
\includegraphics[width=8cm]{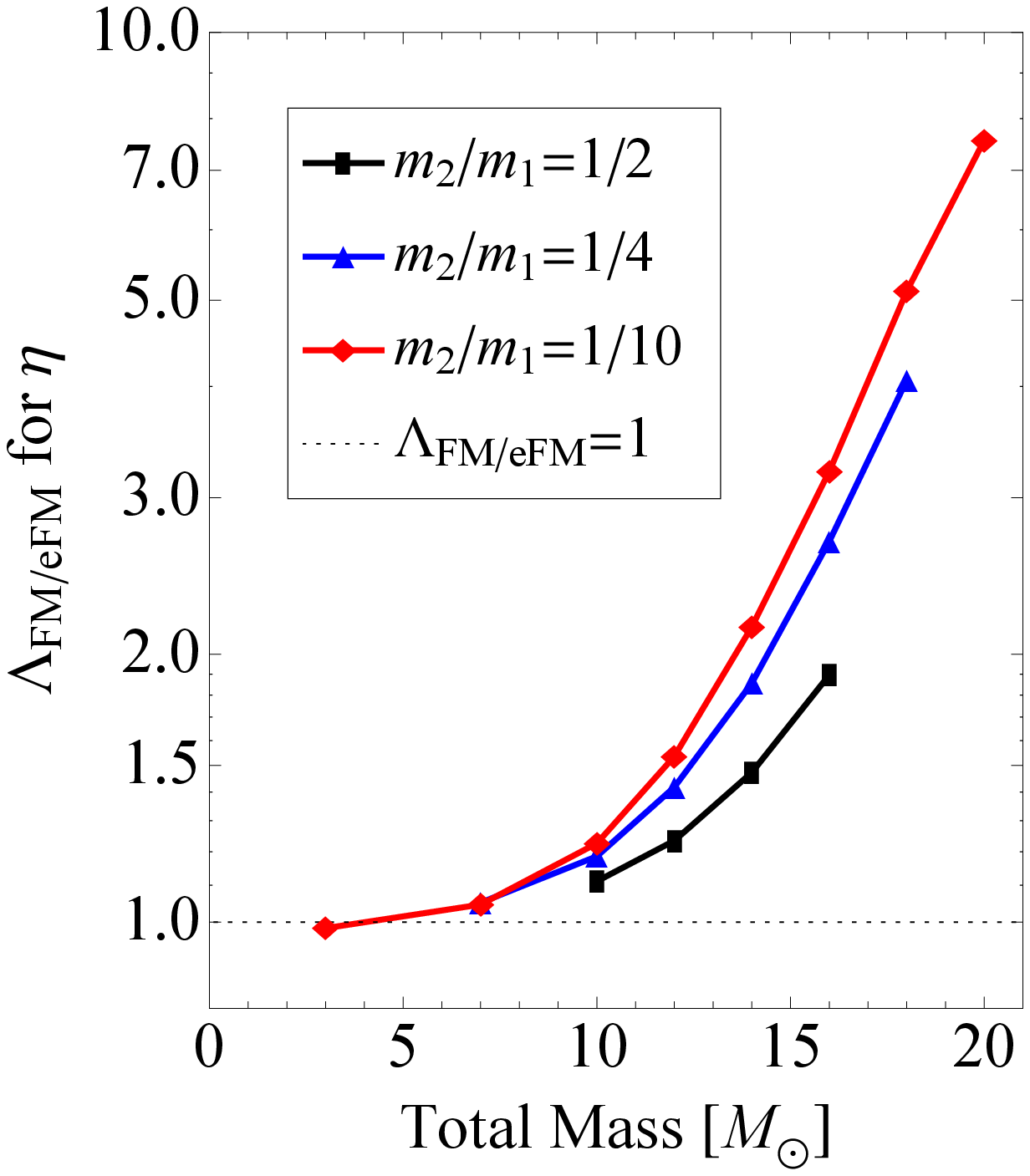}
\caption{\label{fig2}The fractional difference $\Lambda_{\rm FM/eFM}$  as a function of total mass. We use the 2 pN TaylorF2. Note $\Lambda_{\rm FM/eFM}$ increases rapidly for $M > 10\msun$, this trend is consistent with results in RFFM~\cite{Rod13}.}
\end{figure}

By applying the eFM method to the overlap $P_{\rm MCMC}$, we explore various nonspinning binary models with the component masses in the range of $1 \msun \leq m_1, m_2\leq15 \msun$ and total masses in $M\leq20 \msun$ as in RFFM.
For three different values of the input mass ratio, $m_2/m_1=\{1/10, 1/4, 1/2\}$ (i.e., $\eta=\{0.09, 0.16, 0.22\}$), we compare the eFM with the FM in the same way as in RFFM, showing the fractional differences between the FM and eFM standard deviations,
\be
\Lambda_{\rm FM/eFM}\equiv {\sigma^{\rm FM} \over \sigma^{\rm eFM}}.
\ee
The results are summarized in figure~\ref{fig2}.
The FM and eFM results are in good agreement with total masses below $10 \msun$ for both $M_c$ and $\eta$.
However, as the total mass increases, the fractional differences also increase monotonically, up to 5 in $M_c$, and 8 in $\eta$ at high masses ($\sim 20 \msun$).
We find that the overall trend in this result is consistent with that in figure1 of RFFM.
Note that, in this figure, we only consider  2 pN waveforms for consistency with the MCMC simulations in RFFM.
Although we did not perform a direct comparison between the eFM and the MCMC, this result indicates that the eFM can dramatically overcome the limitations of the standard FM in high mass region by  exploring the MCMC posterior surface.
In this work, we only considered sufficiently asymmetric binaries with $m_2 / m_1 \leq 1/2$, because in RFFM the majority of their signals (i.e., 65 $\%$ over 200 simulations) were selected with sufficiently asymmetric mass ratio such that  the 1-$\sigma$ surface about the injected values returned by the FM did not exceed the physical boundary $\eta=0.25$. While RFFM did not give any information on the $\eta$-dependence of $\Lambda_{\rm FM/eFM}$, our result shows that the divergent trend of $\Lambda_{\rm FM/eFM}$ is more pronounced for more symmetric binaries, which will be investigated in detail in the following subsection.

\subsection{Direct comparison to the overlap}
In order to obtain the quantitative comparison between the FM (eFM) and MCMC error estimates,
we calculate the 1-$\sigma$ error of  the MCMC posterior directly from the overlap surface.
From equation~(\ref{eq.P}), the 1-$\sigma$ error for each parameter can be determined by each one-dimensional overlap distribution ($P_{\rm MCMC-1D}$) that is marginalized from the two-dimensional overlap surface $P_{\rm MCMC}$.
For a SNR of 15, we calculate the 68 $\%$ confidence interval ($\sigma^{\rm CI}$) of a parameter $\lambda$ by $\sigma^{\rm CI}_{\lambda}=\delta \lambda=| \lambda_s - \lambda_t |$ where the parameter value of the template ($\lambda_t$) satisfies
\be
P_{\rm MCMC-1D}(\lambda_t)= 1-{0.49 \over 15^2}\simeq0.9978.  \label{eq.P1d}
\ee
Next, we define new fractional differences by
\bea
\Lambda_{\rm FM/CI }&\equiv&  {\sigma^{\rm FM} \over \sigma^{\rm CI}}, \\
\Lambda_{\rm eFM/CI }&\equiv&  {\sigma^{\rm eFM} \over \sigma^{\rm CI}},
\eea
where, since the eFM is computed by the two-dimensional overlap surface, the fitting region is
 $P_{\rm min}=0.995$, and the $\sigma^{\rm eFM}$ should be obtained by the  $2 \times 2$ eFM.

\begin{figure}[t]
\includegraphics[width=14cm]{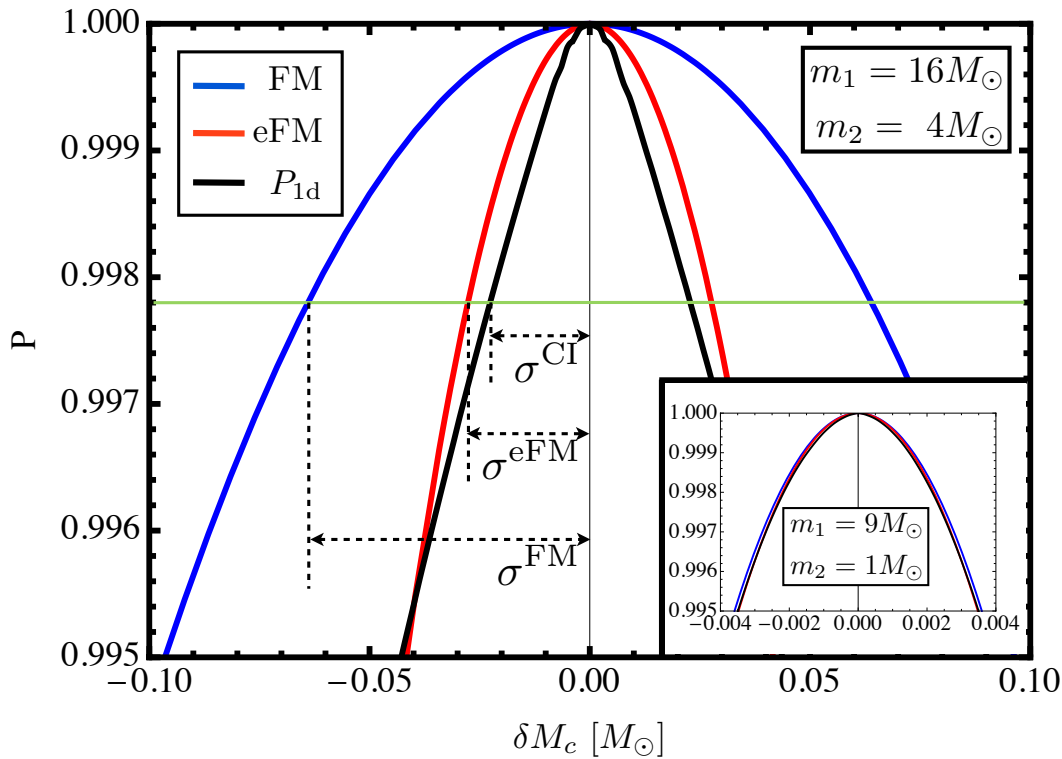}
\caption{\label{fig3} Schematic view showing how to calculate the $\sigma^{\rm CI}$ and direct comparisons between three methods. The blue and red lines are quadratic functions calculated by the $\sigma^{\rm FM}$ and  $\sigma^{\rm eFM}$, respectively. The black line is the one-dimensional overlap distribution ($P_{\rm MCMC-1D}$) marginalized from the two-dimensional overlap surface $P_{\rm MCMC}$. The green line indicates $P=0.9978$ [see, equation~(\ref{eq.P1d})]. For comparison, the small plot is given in the bottom right with smaller component masses, note that the three lines are almost coincide.}
\end{figure}

In figure~\ref{fig3}, we describe how to calculate the $\sigma^{\rm CI}$ from the $P_{\rm MCMC-1D}$ showing direct comparisons to the others.
Here, while the $\sigma^{\rm CI}$ is determined by the $P_{\rm MCMC-1D}$, the blue and red lines are quadratic functions oppositely derived by the
$\sigma^{\rm FM}$ and  $\sigma^{\rm eFM}$, respectively.
As discussed in \cite{Man14}, the overlap in the MCMC formalism can have a sharp peak at the origin for a high mass system, then the fitting function may not be valid at a very small scale. However, at a sufficiently large fitting scale, the overall surface can be well approximated by a quadratic function (e.g., see, \cite{Wad13,Cho13}). Therefore, the accuracy of the eFM method directly depends on the scale of the fitting region, which will be discussed in figure~\ref{fig5} in more detail.

\begin{figure}[t]
\includegraphics[width=\columnwidth]{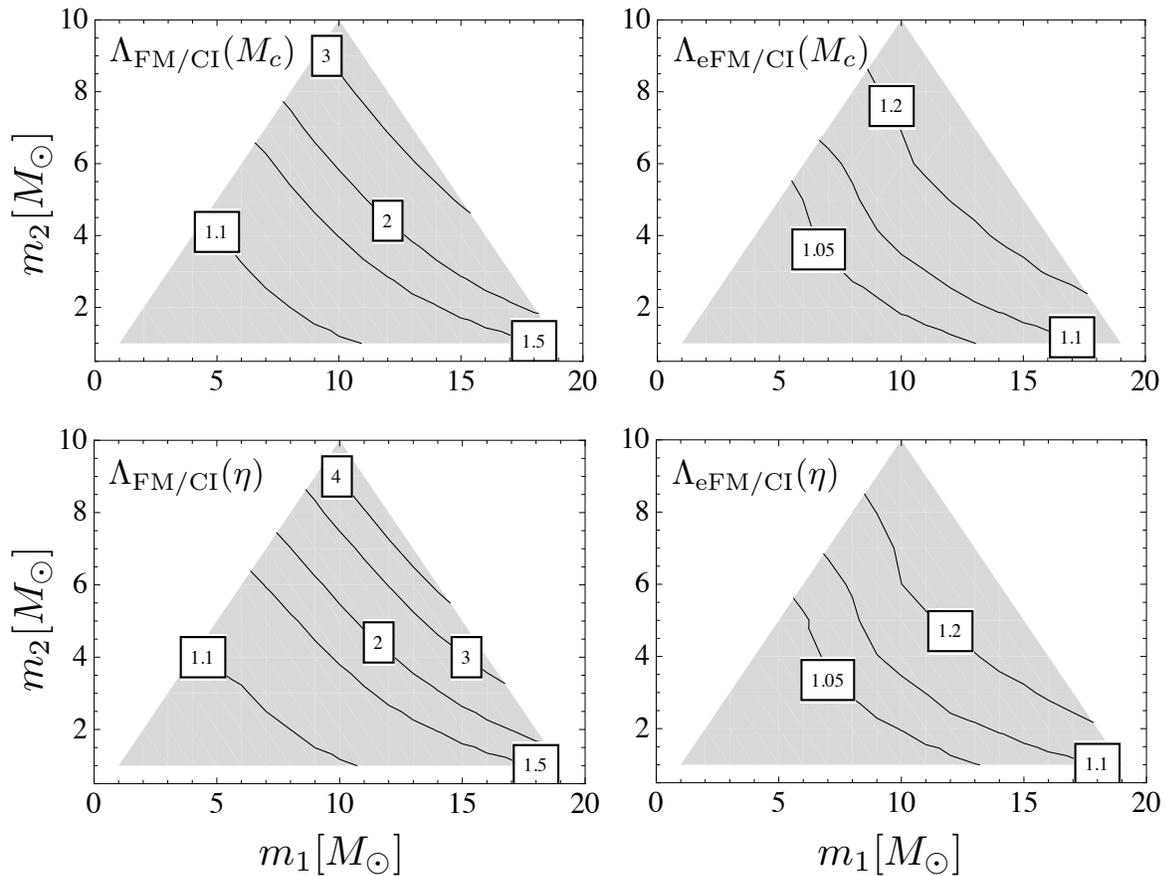}
\caption{\label{fig4}The contour plot of $\Lambda_{\rm FM/CI }$ and $\Lambda_{\rm eFM/CI }$. Note that $\Lambda_{\rm FM/CI }$ increases up to 3 or 4 as the total mass increases, while $\Lambda_{\rm eFM/CI }$ remains below 1.25 for all mass regions.}
\end{figure}

To investigate the acceptable criteria of the FM and eFM for estimation of the MCMC error bounds, we summarize $\Lambda_{\rm FM/CI }$ and $\Lambda_{\rm eFM/CI }$ for all mass regions in figure~\ref{fig4}, which is the main result of this work.
As in figure~\ref{fig2}, one can find that $\Lambda_{\rm FM/CI }$ strongly depends on the total mass for both mass parameters, and increases up to 3 or 4. In addition, one can also find a weak dependence on the mass ratio.
On the other hand, though the overall shape of the contours is similar to that of $\Lambda_{\rm FM/CI }$, $\Lambda_{\rm eFM/CI}$ is considerably suppressed. We find that the eFM can approximate the MCMC error bounds with an accuracy of $\Lambda_{\rm eFM/CI}\simeq1.2$ at high masses ($\sim 20 \msun$).

\subsection{Dependence on the signal strength (SNR)}
Mandel {\it et al.}~\cite{Man14} demonstrated that the the difference between the FM and the MCMC can also be dependent on the SNR as well as the total mass.
In figures~\ref{fig1} and \ref{fig3}, one can deduce that for a given SNR or a fitting region, a sharpness of the overlap surface $P_{\rm MCMC}$ is more significant as the system mass increases, giving the larger difference from the  $P_{\rm FM}$. This behavior indeed results in the divergent trend of the fractional differences in figures~\ref{fig2} and \ref{fig4}.
In the same manner, for a given binary mass, if we select a smaller fitting region, the sharpness can be more significant. Therefore, the fractional difference can be larger for the higher SNR.

\begin{figure}[t]
\includegraphics[width=8cm]{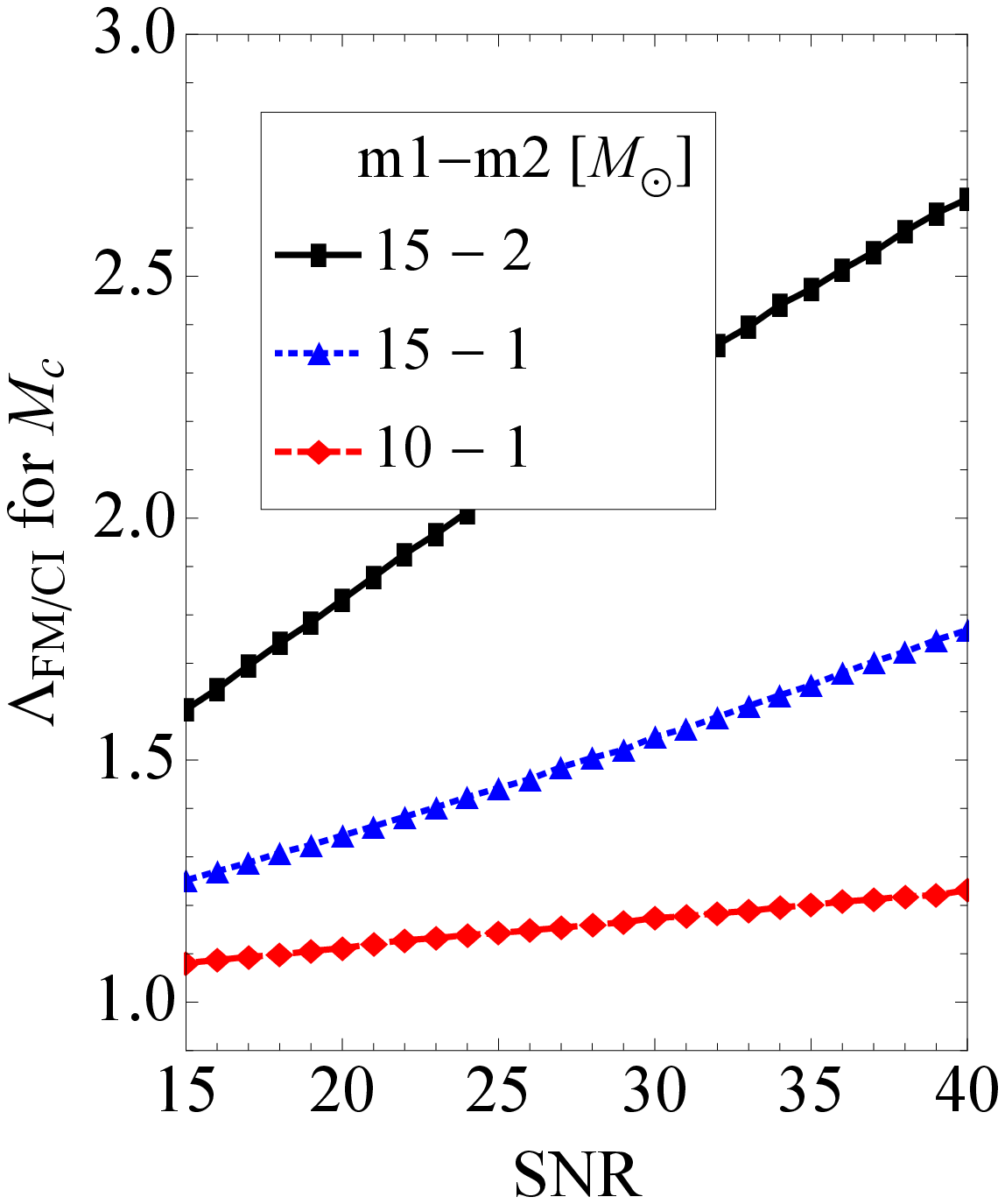}
\includegraphics[width=8cm]{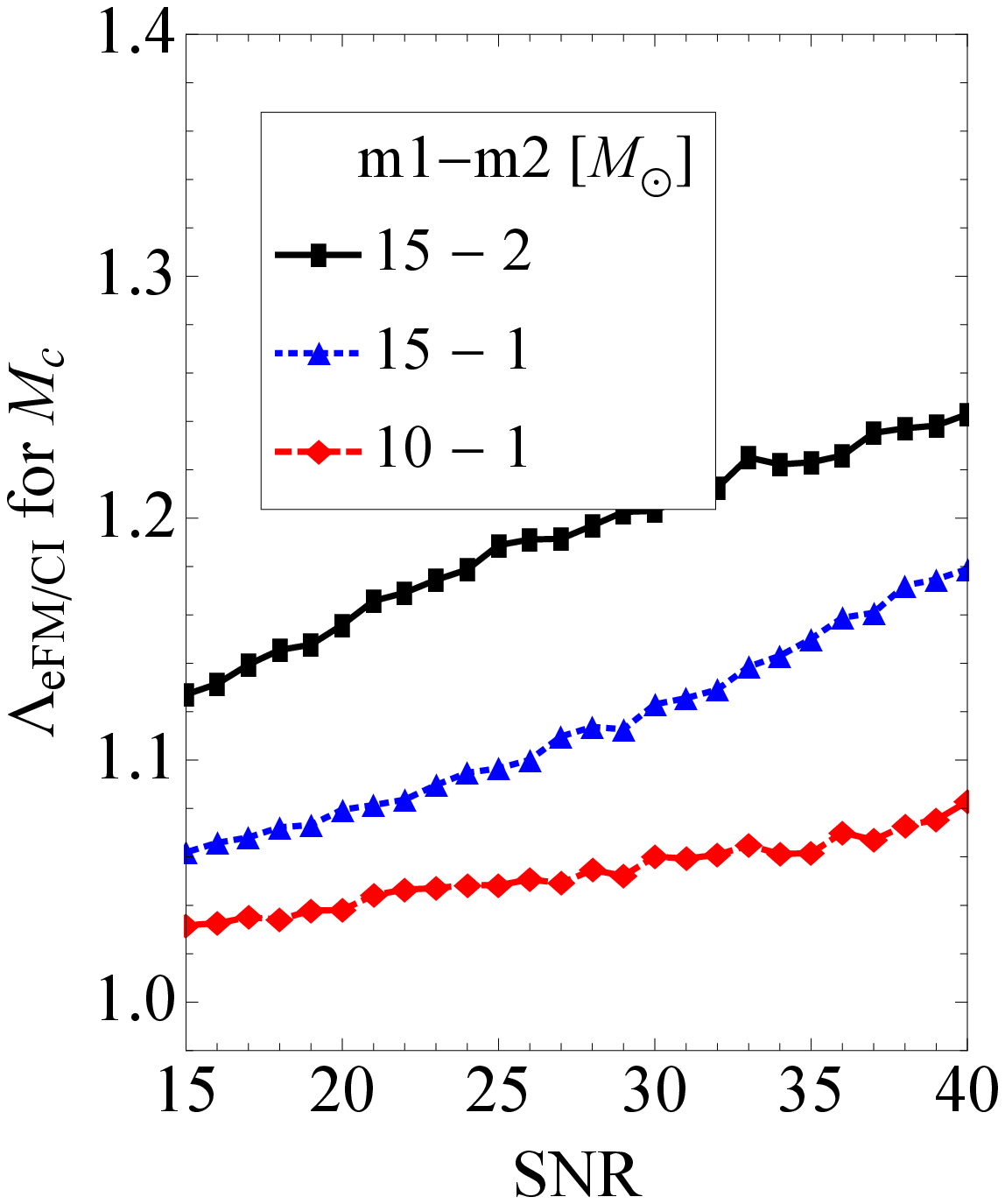}
\caption{\label{fig5}The fractional difference $\Lambda_{\rm FM/CI }$ and $\Lambda_{\rm eFM/CI }$ as a function of SNR.  For the eFM, the fitting region is related with the SNR as in equation~(\ref{eq.P2d}).
Both $\Lambda_{\rm FM/CI}$ and  $\Lambda_{\rm eFM/CI}$  increase with SNR because the sharpness of the overlap surface becomes more pronounced as  $P_{\rm min}$ approaches unity. However, note that the rates of increase of $\Lambda_{\rm eFM/CI}$ are considerably smaller than those of $\Lambda_{\rm FM/CI}$.}
\end{figure}

In figure~\ref{fig5}, we show the dependence of the fractional errors on the SNR.
In the left panel, $\Lambda_{\rm FM/CI }$ increases with SNR for all binary models given.
This implies that the validity of the FM can be broken even for the low mass systems if the SNR is sufficiently high. 
In this case, the contours in figure~\ref{fig4} will be shifted towards the low mass region for a higher SNR.
In the right panel, $\Lambda_{\rm eFM/CI}$ also increases with SNR due to the similar reason to the FM case.
As the sharpness of $P_{\rm MCMC}$ becomes more significant, the accuracy of the quadratic fitting function decreases.
However, the rates of increase of $\Lambda_{\rm eFM/CI}$ are considerably smaller than those of $\Lambda_{\rm FM/CI}$, and this indicates that the eFM is acceptable even for very strong signals.

It should be mentioned that the SNR dependence in this result is counter intuitive to the common understanding (i.e., the FM gives better results for high SNR cases). However, this is because
the unphysical $f_{\rm cut}$ of the TaylorF2 model can introduce artificial structures in the MCMC posterior distribution, that may cause the sharpness at the origin. So, the MCMC result cannot be accurate at the very high SNR for the inspiral-only waveforms. The impact on the overall posterior distribution has not been studied yet. In this work, we do not take into account the full effect caused by the artificial cutoff $f_{\rm cut}$, which is beyond the scope of this work.

\section{Summary and discussion}\label{sec.conclusion}

In this work, we demonstrated the inadequacy of the standard FM for the TaylorF2 waveform, especially for the high mass region.
In spite of the considerable computational advantages of the TaylorF2 model~\cite{Lvc13}, the FM is consistent with the MCMC only at a mass range of $M<10 \msun$ for the initial LIGO sensitivity, and the mass range drops off quickly for the higher SNR signal.
By comparing the standard FM and MCMC overlap formalisms, we showed the origin of the discrepancy between the FM and MCMC error predictions summarized in RFFM.

We briefly reviewed the eFM approach and discussed the validity of the effective method.
By applying the eFM method to the MCMC posteriors ($P_{\rm MCMC}$), we found  that the overall trend of the fractional difference between the FM and eFM errors is consistent with that of RFFM, and the eFM method can be more acceptable than the standard FM in the estimation of the MCMC error bounds for total masses  $M \leq 20 \msun$. 
In addition, the accuracy of the eFM weakly depends on the SNR, so the eFM is broadly acceptable even for very strong signals.

In this work, we applied the eFM method only to the frequency domain inspiral waveform model. 
However, as mentioned above, the MCMC posterior can have a limitation itself for high SNR cases due to the unphysical $f_{\rm cut}$ of the waveforms.
The accuracy of the MCMC and eFM results for the TaylorF2 model at a given SNR should be further investigated by comparing with the full MCMC result which incorporates the full inspiral-merger-ringdown waveforms,
and our method can be easily applied to the full waveforms.
In addition,  since it is not necessary to directly differentiate the wave function in the eFM formalism, this method can also be applied to other complicated waveform models, for example, the time domain waveforms.
Recently, O'Shaughnessy {\it et al.}~\cite{Osh141,Osh142} showed a good agreement between the eFM and the MCMC using a time domain inspiral waveform for a black hole-neutron star binary with masses of $10 \msun$ and $1.4 \msun$.

Since the overlap computation in the FM formalism involves a detector noise spectrum, our result explicitly depends on the detector characteristics.
Although we adopted the initial LIGO noise spectrum in this work, the Advanced LIGO noise curve~\cite{Lig10} should be taken into account in future studies for more realistic results, and this work can be easily extended to other detector noise spectrums.


%

\ack{The authors would like to thank Richard O'Shaugnessy and anonymous LIGO Scientific Collaboration reviewers for helpful comments.
The authors also thank the referees for their valuable comments and suggestions.
This study was financially supported by the 2013 Post-Doc. Development Program of Pusan National University. H. S. C. and C. H. L. are supported in part by the National Research Foundation Grant funded by the Korean Government (No. NRF-2011-220-C00029) and the BAERI Nuclear R \& D program (No. M20808740002) of Korea. This work uses computing resources at the KISTI GSDC.}

%
%
%

\end{document}